\documentstyle[sprocl,epsfig]{article}

\def\NPB{{\em Nucl. Phys.} B}
\def\PLB{{\em Phys. Lett.}  B}

\def\PRD{{\em Phys. Rev.} D}

\def\JMP{{\em J. Math. Phys.} }

\def\lsim{\mathrel{\raise.3ex\hbox{$<$\kern-.75em\lower1ex\hbox{$\sim$}}}}
\def\lg{\mathrel{\raise.3ex\hbox{$<$\kern-.90em\lower1ex\hbox{$>$}}}}

\def\om{\omega}
\def\Ga{\Gamma}

\def\be{\begin{equation}}
\def\ee{\end{equation}}
\def\bea{\begin{eqnarray}}
\def\eea{\end{eqnarray}}

\begin{document}

\title{READING THE NUMBER OF EXTRA DIMENSIONS IN THE SPECTRUM OF HAWKING
RADIATION}

\author{P. KANTI}

\address{Theory Division, CERN, CH-1211 Geneva 23, Switzerland\\
E-mail: panagiota.kanti@cern.ch} 

\maketitle\abstracts{ After a brief review of the production and decay of 
Schwarzschild-like $(4+n)$-dimensional black holes in the framework of theories
with Large Extra Dimensions, we proceed to derive the greybody factors and
emission rates for scalars, fermions and gauge bosons on the brane. We
present and discuss analytic and numerical methods for obtaining the
above results, and demonstrate that both the {\it amount}
and {\it type} of Hawking radiation emitted by the black hole can help
us to determine the number of spacelike dimensions that exist in nature.}

\section{Introduction}

The idea of the existence of {\it extra spacelike} dimensions in nature 
has been revived during the last few years in an
attempt to explain the gap between different energy scales \cite{hierarchy}.
It demands the introduction of (at least) one 3-brane,
which plays the role of our 4-dimensional world on which all Standard
Model fields are localized; on the other hand, gravity can propagate both
on the brane and in the bulk -- the spacetime transverse to the brane. 
The introduction of extra spacelike dimensions inevitably affects both
gravitational interactions and particle physics phenomenology and may 
lead to modifications in standard cosmology \cite{cosmo}.
In this talk, we will focus on the scenario that postulates the existence of
{\it Large Extra Dimensions} \cite{ADD}, according to which, in nature,
there are $n \geq 2$ extra, spacelike, compact dimensions with size
$R \leq 1$\,mm leading to
\begin{equation}
M_P^2 \sim M_*^{2+n}\,R^n\,.
\end{equation}
From the above, it becomes clear that if $R \gg l_P$, then the 
$(4+n)$-dimensional Planck mass, $M_*$, will be much lower than the
4-dimensional one, $M_P$. 

The above scenario can easily accommodate the notion of the existence of
Black Holes in the universe. These objects may appear in nature 
with a variety of sizes, 
therefore, a black hole with 
$l_P< r_H < R$ is clearly a higher-dimensional
object that can be still treated semi-clasically. These {\it mini} black
holes are centered on the brane because the ordinary matter that undergoes 
gravitational collapse is restricted to live on the brane. Such mini
black holes may have been created in the early universe due to density
perturbations and phase transitions. Recently, there was a novel proposal
according to which small, higher-dimensional black holes may be created
during scattering of highly energetic particles at colliders or at the
earth's atmosphere \cite{BF,GT,DL}. This may happen when any two partons of the
colliding particles pass within the horizon radius corresponding to their
centre-of-mass energy \cite{thorne}. The created black holes will go through
a number of stages in their life \cite{GT}, namely: {\bf (i)} the 
{\it balding phase}, where the black hole sheds the {\it hair} inherited from
the original particles, and the asymmetry due to the violent production
process (15\% of the total energy); {\bf (ii)} the {\it spin-down phase},
in which the black hole looses its angular momentum through the emission of
Hawking radiation \cite{hawking} and, possibly, through superradiance 
(25\% of the total energy); {\bf (iii)} the {\it Schwarzschild phase}, where
the emission of Hawking radiation results in the decrease of its mass (60\%
of the total energy); and, {\bf (iv)} the {\it Planck phase}, that starts when
$M_{BH} \sim M_*$ and whose study demands a theory of quantum gravity.

In this work, we concentrate on the spherically-symmetric Schwarzschild phase
of a $(4+n)$-dimensional black hole, which is the longer one and
accounts for the greatest proportion of the mass loss through the emission
of Hawking radiation. The spacetime around such a black hole is given
by \cite{MP}
\begin{equation}
ds^2 = - \left[1-\left(\frac{r_H}{r}\right)^{n+1}\right]\,dt^2 +
\left[1-\left(\frac{r_H}{r}\right)^{n+1}\right]^{-1}\,dr^2 + 
r^2 d\Omega_{2+n}^2\,,
\label{metric-n}
\end{equation}
where
\begin{equation}
d\Omega_{2+n}^2=d\theta^2_{n+1} + \sin^2\theta_{n+1} \,\biggl(d\theta_n^2 +
\sin^2\theta_n\,\Bigl(\,... + \sin^2\theta_2\,(d\theta_1^2 + \sin^2 \theta_1
\,d\varphi^2)\,...\,\Bigr)\biggr)\,.
\end{equation}
In the above, $0 <\varphi < 2 \pi$ and $0< \theta_i < \pi$, for 
$i=1, ..., n+1$. The horizon radius and temperature of such a black hole
are given by
\begin{equation}
r_H= \frac{1}{\sqrt{\pi}M_*}\left(\frac{M_{BH}}{M_*}\right)^
{\frac{1}{n+1}}\left(\frac{8\Gamma\left(\frac{n+3}{2}\right)}{n+2}\right)
^{\frac{1}{n+1}}\,, \quad T_{BH} = {(1 + n) \over 4\pi\,r_H}\,,
\end{equation}
respectively, where $M_{BH}$ is the mass of the black hole.

A black hole with temperature $T_{BH}$ emits Hawking
radiation\,\footnote{The Hawking radiation can be conceived as the creation
of a virtual pair of particles just outside the horizon -- the antiparticle
falls into the BH, the particle escapes to infinity.} with an 
{\it almost} blackbody spectrum and an energy emission rate given by
\begin{equation}
\frac{dE(\om)}{dt} = \sum_{j} {\sigma_{j}(\om)\,\,\om  \over
\exp\left(\om/T_{BH}\right) \mp 1}\,\,\frac{d^{n+3}k}{(2\pi)^{n+3}}\,.
\label{rate}
\end{equation}
In the above, $\omega$ is the energy of the emitted particle, and the
statistics factor in the denominator is $-1$ for bosons and $+1$ for fermions.
The $\omega$-dependent factor $\sigma_{j}$, with $j$ being the total angular
momentum number, is the so-called {\it greybody
factor} that distorts the blackbody spectrum; this is due to the fact that
any particle emitted by the black hole has to traverse a strong gravitational
background before reaching the observer, and $\sigma_{j}$ stands for the
corresponding transmission cross-section. If, for example, a scalar field 
of the form $\phi\,(t,r,\theta_i,\varphi)=
e^{-i\om t}\,R_{\om j}(r)\,{Y}_j(\Omega)$ propagates in the
background of Eq. (\ref{metric-n}), its radial equation of motion, in terms
of the `tortoise' coordinate $y=\frac{\ln h(r)}{r_H^{n+1}\,(n+1)}$, with
$h(r) = 1-(\frac{r_H}{r})^{n+1}$, reads
\begin{equation}
\left(-\frac{d^2 \,}{dy^2} + r^{2n+4}\left[-\omega^2
+\frac{j(j+n+1) h(r)}{r^2}\right] \right) R(y) = 0\,.
\end{equation}
The quantity inside square brackets gives the potential barrier in
the $r>r_H$ area, and its expression reveals the fact that all parameters
$(\omega, j, n)$ affect the
value of $\sigma_j$. The latter is formally defined as \cite{GKT}
\begin{equation}
\sigma_j(\omega) = \frac{2^{n}\pi^{(n+1)/2}\,\Gamma[(n+1)/2]}{ n!\,
\omega^{n+2}}\,\frac{(2j+n+1)\,(j+n)!}{j !}\,|{\cal A}_j|^2\,,
\label{grey-n}
\end{equation}
where ${\cal A}_j$ is the corresponding Absorption Probability
for a particle moving towards the black hole.

The $(4+n)$-dimensional black hole (\ref{metric-n}) emits Hawking radiation
both in the bulk and on the brane. Under the assumptions of the theory,
only gravitons, and possibly scalar fields, live in the bulk and 
therefore these are the only particles that can be emitted in the bulk. 
On the other hand, the brane-localised modes include zero-mode scalars,
fermions, gauge bosons and zero-mode gravitons. It is therefore much
more interesting, from the observational point of view, to study the
emission of brane-localised modes. The 4-dimensional background in which
these modes propagate is the projection of the line-element
(\ref{metric-n}) onto the brane, which is realized by setting
$\theta_i=\pi/2$, for $i>1$. Then, we obtain
\begin{equation}
ds^2_4 = - \left[1-\left(\frac{r_H}{r}\right)^{n+1}\right]\,dt^2 +
\left[1-\left(\frac{r_H}{r}\right)^{n+1}\right]^{-1}\,dr^2 + 
r^2\,(d\theta^2 + \sin^2\theta\,d\varphi^2)\,.
\label{metric-4}
\end{equation}
The emission of brane-localised modes is a 4-dimensional process and the
corresponding greybody factor takes the simplified form
\begin{equation}
\sigma_j(\omega) = \frac{\pi}
{ \omega^{2}}\,(2j+1)\,|{\cal A}_j|^2\,.
\end{equation}
Nevertheless, ${\cal A}_j$ {\it does} depend on the number $n$ of extra
dimensions projected onto the brane, through the expression of the line-element
(\ref{metric-4}), and this valuable piece of information will be evident in the
computed radiation spectrum.

\section{Hawking radiation from a Black Hole on a 4-dimensional brane}

In these section, we will briefly discuss the emission of brane-localised
scalars, fermions and gauge bosons (for more information, see Refs.
\cite{kmr1,kmr2,hk}). We will present analytic and numerical methods for
deriving the desired results, and comment on the dependence of the amount
and type of particles emitted on the brane on the number of extra dimensions.

The derivation of a {\it master equation} \cite{hk} describing the motion 
of a particle with arbitrary spin $s$ in the projected background
(\ref{metric-4}) is the starting point of our analysis. By using the factorization
\begin{equation}
\Psi_s=e^{-i\omega t}\,e^{im\varphi}\,R_s(r)\,S_{s,j}^m(\theta)\,,
\end{equation}
where $Y_{s,j}^m=e^{im\varphi}\,S_{s,j}^m(\theta)$ are the spin-weighted
spherical harmonics \cite{goldberg}, and by employing the Newman-Penrose method,
we obtain the radial equation
\begin{equation}
\label{radial}
\frac{1}{\Delta^s} \frac{d}{dr}\left(\Delta^{s+1} \frac{d R_s}{dr}\right)+
\left[\frac{\omega^2 r^2}{h}+2i\omega s r-\frac{is\omega r^2 h'}{h}+
s(\Delta''-2)-\lambda_{sj}\right] R_s (r)=0,
\end{equation}
where $\Delta=hr^2$ and $\lambda_{sj}=j(j+1)-s(s+1)$.

The above equation needs to be solved over the whole radial domain. This can
be done either analytically or numerically with different advantages in each
case. The analytic treatment demands the use of an approximate method in which
the above equation is solved at the near-horizon regime ($r \simeq r_H$) and
far-field regime ($r \gg r_H$), and the corresponding solutions are matched
at an intermediate zone. This method allows us to solve the radial equation 
for particles with arbitrary spin $s$ in a unified way. This was done in
Ref. \cite{kmr2}, where the transformed, and slightly simpler, equation
\begin{equation}
\label{radial1}
\Delta^s \frac{d}{dr}\left(\Delta^{1-s} \frac{d P_s}{dr}\right)+
\left[\frac{\omega^2 r^2}{h}+2i\omega s r-\frac{is\omega r^2 h'}{h}
-\Lambda_{sj}\right] P_s (r)=0
\end{equation}
was used, where now $\Lambda_{sj}=j(j+1)-s(s-1)$ and $P_s(r)=\Delta^s R_s(r)$.
At the near-horizon regime, a change of variable $ r \rightarrow h(r)$ brings
the above differential equation to a hypergeometric form. Under the boundary
condition that only incoming modes must exist close to the horizon, its general
solution reduces to
\begin{equation}
P_{NH}(h)=A_h \,h^{\alpha}\,(1-h)^\beta\,F(a,b,c;h)\,,
\end{equation}
where $a=\alpha + \beta +\frac{s + n\,(1-s)}{(n+1)}$,  
$b=\alpha + \beta$, and $c=1-s + 2 \alpha$, with
\begin{equation}
\alpha = -\frac{i \om r_H}{n+1}, \quad
\beta =\frac{1}{2(n+1)}\,\biggl[\,1-2s - \sqrt{(1+2j)^2 -
4 \om^2 r_H^2 -8 is \om r_H}\,\,\biggr],
\end{equation}
and $A_h$ an arbitrary constant. At the far-field regime, on the other hand,
by setting $r \gg r_H$ in Eq. (\ref{radial1}), we obtain a confluent
hypergeometric equation with solution:
\begin{equation}
P_{FF}(r) = e^{-i \om r} r^{j+s} \Bigl[B_+\,M(j-s+1,2j+2, 2i\om r) 
+ B_-\,U(j-s+1, 2j+2, 2 i \om r)\Bigr]\,,
\label{FF-sol}
\end{equation}
where $M$ and $U$ are the Kummer functions and $B_\pm$ arbitrary coefficients.
We finally match the two asymptotic solutions: we first expand
the NH solution in the limit $ r \gg r_H $, and the FF solution in the limit
$ \omega r \ll 1$. Both solutions then take the form 
$P = C_1 r^j + C_2\,r^{-(j +1)}$, and the matching of the corresponding 
coefficients determines the ratios $B_\pm/A_h$. These constant ratios are
involved in the definition of the Absorption Probability ${\cal A}_j$, 
that is defined as:
\begin{equation}
|{\cal A}_j|^2= \frac{{\cal F}_{in}^{(h)}}{{\cal F}_{in}^{(\infty)}}\,,
\label{absorption}
\end{equation}
in terms of the energy fluxes evaluated at infinity and the horizon. The ratios
$B_\pm/A_h$ are given by complicated expressions \cite{kmr1,kmr2} that depend on
$\omega$, $r_H$, $j$ and $n$, and their behaviour, as any of these parameters
change, is difficult to infer. 

A usual manipulation is to expand these complicated expressions in the limit
$\om r_H \ll 1$. Then, by keeping the dominant term in the power series, we obtain
simple, elegant expressions for ${\cal A}_j$; for example, in the case of
scalar fields, this takes the form \cite{kmr1}
\begin{equation}
|{\cal A}_j|^2 = \frac{16 \pi}{(n+1)^2}\,
\biggl(\frac{\om r_H}{2}\biggl)^{2j+2}\,
\frac{\Ga(\frac{j+1}{n+1})^2\,\Ga(1+\frac{j}{n+1})^2}
{\Ga(\frac{1}{2}+j)^2\,\Ga(1 +\frac{2j+1}{n+1})^2} + ... \,.
\label{simple}
\end{equation}
According to the above, $|{\cal A}_j|^2$, and subsequently the greybody factor,
decreases as $j$ increases while it gets enhanced as $n$ increases. Similar
expressions and results are obtained for the emission of fermions and gauge fields. 

Before using the {\it simplified expressions} for the evaluation of the
emission rates from the black hole, let us check their validity. These
expressions have been derived in the low-energy limit and thus we expect
them to break down at intermediate values of $\omega r_H$. But can we trust
them even in the low-energy regime? The 
{\it full analytic expressions} \cite{kmr1,kmr2} (the ones derived before
expanding in power series in $\omega r_H$) can provide the answer. Although
more complicated,
these analytic expressions can be easily plotted to reveal the behavior in
question. A simple analysis shows that, in the case of scalar fields, the
greybody factor indeed decreases with $j$ but it also decreases with $n$, the
latter being in apparent contradiction with Eq. (\ref{simple}).
A similar analysis for
the case of fermions and gauge fields leads to a satisfactory agreement with
the simplified analytic expressions. So, what went wrong with the scalar fields?
The main contribution to the greybody factor for spin-0 fields comes from 
the dominant first partial wave ($j=0$) in Eq. (\ref{simple}), which is found
to be independent of $n$. Any dependence on $n$ must be therefore read from the
higher partial waves which indeed are enhanced as $n$ increases. However, it turns
out that the next-to-leading order terms, denoted by ``..." in Eq. (\ref{simple}),
are of the same order as the higher partial waves and should also be taken into
account. When this is done, we are finally led to a decrease with $n$. This
does not happen for fields with $s={1 \over 2}, 1$ where the lowest partial wave
has a non-trivial dependence on $n$.

Going back to the {\it full analytic expressions} for the absorption probability,
we may safely use them to derive the greybody factors and emission rates since
they include all subdominant corrections and thus have an extended validity
up to higher energies. As mentioned above, the greybody factor for scalar fields
is found \cite{kmr2} to decrease with $n$, while the ones for fermions and gauge
bosons are enhanced as $n$ increases, at least up to intermediate energies.
While scalars and fermions have a non-vanishing greybody factor as 
$\omega r_H \rightarrow 0$, the one for the gauge bosons vanishes.
By using Eq.~(\ref{rate}) with $n=0$, we may then compute the emission rates
for different particles on the brane \cite{kmr2}. Despite the different behaviour
of the greybody factors for scalars, fermions and gauge bosons, the corresponding
emission rates exhibit a universal behaviour according to which the
energy emitted per unit time and energy interval is strongly enhanced as $n$
increases up to intermediate values of the energy parameter $\omega r_H$.

But what about the high-energy behaviour? The full analytic expressions for the
absorption coefficient, although of extended validity compared to the simplified
ones, are not valid in the high-energy regime. The assumption that
$\omega r_H \ll 1$ was made in the matching process, therefore, these
expressions are also bound to break down at some point. The only way to
derive exact results valid at all energy regimes is through numerical
integration. This task was performed in Ref.\,\cite{hk} where both greybody
factors and emission rates were computed. The {\it exact results} for the greybody
factors closely follow the ones predicted by the full analytic results at
the low- and intermediate-energy regimes. At the high-energy regime, a universal
asymptotic behaviour, the same for all particle species, is revealed: the black
hole acts as an absorptive area of radius\,\cite{sanchez} $r_c$ that leads
to a blackbody spectrum with $\sigma \simeq \pi r_c^2$. The effective
radius $r_c$ is $n$-dependent and decreases as the value of $n$
increases\,\cite{Emparan}. 

The energy emission rates determined by the numerical analysis also agree
with the ones derived by using the full analytic expressions up to intermediate
energies. Whereas the analytic results covered only a part of the ``greybody" curve,
especially as the spin of the particle and the number of extra dimensions increased,
the numerical results allow us to derive the complete radiation spectrum. In Fig.1,
we depict the energy  emission rates for fermions on the brane. The
enhancement of the rate, at which energy is radiated on the brane, as $n$ 
increases, is indeed impressive reaching orders of magnitude for large values
of $n$, and for all species of particles. This is mainly due to the increase in the
temperature of the black hole that automatically increases the amount of energy
that can be spent on the emission of particle modes. Apart from the increase in
the height and width of each curve, a Wien's type of displacement of the peak
towards higher energies also takes place. 
\begin{figure}[t]
\centerline{\psfig{file=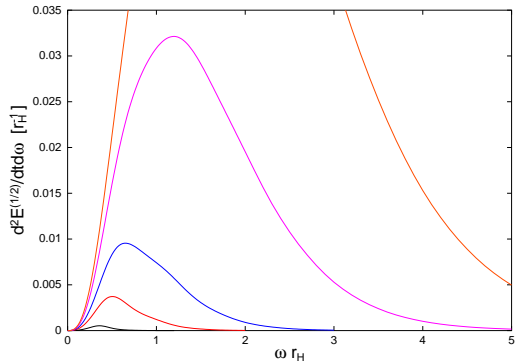,height=1.9in}}
\caption{Energy emission rates for fermions on the brane by a
Schwarzschild-like (4+n)D black hole. The lines correspond to $n=0,1,2,4$ 
and 6, respectively, increasing upwards.}
\end{figure}

The magnitude of the enhancement of the emission rates is different for different
species of particles. This evidently changes the type of particle that the
black hole prefers to emit, as the number of extra dimensions changes. In 4
dimensions\,\cite{page,Webber}, most of the energy of the black hole is spent on
the emission of scalar degrees of freedom, with the emission of fermions and
gauge degrees \cite{generators} being subdominant. Our analysis\,\cite{hk}
reveals that the latter
two species take only 0.55 and 0.23, respectively, of the energy spent on the
scalar ``channel". As $n$ increases, however, these numbers drastically change:
for $n=2$, the relative emissivities become 1\,:\,0.91\,:\,0.91, and, for $n=6$,
1\,:\,0.84\,:\,1.06. Therefore, for
intermediate values of $n$, the black hole spends the same amount of energy for
the emission of fermions and gauge bosons, while, for large values of $n$, the
emission of gauge fields is the dominant ``channel". 

In conclusion, the detection of Hawking radiation from a decaying black hole
is an exciting challenge on its own. If Extra Dimensions exist, the emission
spectra can also help us to determine the {\it dimensionality} of spacetime, even
if we look only at the energy emitted on the brane 
(for exact results on the Bulk-to-Brane relative emissivities, see \cite{hk}). 
Distinctive features of the $n$-dependent spectrum is the {\it amount} and 
{\it type} of radiation emitted by the black hole, and the improvement
of the spectrum by implementing the exact greybody factors is imperative
in order to accomplish this task in an accurate way.

\section*{Acknowledgments}
I would like to thank John March-Russell and Chris M. Harris for a
constru\-ctive and enjoyable collaboration. This contribution to the
2003 String Phenomeno\-logy Conference proceedings is dedicated to the
memory of Ian I. Kogan, one of the most supportive and stimulating
collaborators I have ever had.

\end{document}